\documentclass{ws-procs9x6}

\def\gev{~{\rm GeV}}
\def\ale{\alpha_0}
\def\als{\alpha_{\rm s}}

\newcommand{\gsim}{\raisebox{-4pt}{$\,\stackrel{\textstyle
                                                         >}{\sim}\,$}}

\begin{document}

\title{The Handbag Contribution to Two-Photon Annihilation 
       Into Meson Pairs}

\author{C. Vogt}

\address{Nordita \\ Blegdamsvej 17 \\ 2100 Copenhagen, Denmark \\
         Email: cvogt@nordita.dk}

\maketitle

\abstracts{We report on the handbag contribution to two-photon 
annihilation into pion and kaon pairs at large energy and momentum 
transfer. The underlying physics of the mechanism is outlined and 
characteristic features and predictions are presented.
(Talk given at workshops ``QCD-N'02'', Ferrara, Italy, April 3-6, 
and ``Exclusive Processes at High Momentum Transfer'', 
Jefferson Lab, Newport News, VA, USA, May 15-18, 2002.)} 

\section{Introduction}

Meson pair production in the collision of two real photons at 
asymptotically large energies can be described in the hard 
scattering approach~\cite{bl1980}, where to leading-twist the 
transition amplitude factorizes into a perturbatively calculable 
$\gamma \gamma \to q\bar{q} \, q\bar{q}$ subprocess and 
single-meson distribution amplitudes for the hadronization 
of each of the $q\bar{q}$ pairs. 
The perturbative contribution to the cross section for 
$\gamma\gamma \to \pi^+ \pi^-$, however, turns out to be well
below the experimental data if single-pion distribution 
amplitudes consistent with other data are employed~\cite{vogt2000}.

\begin{figure}[th]
\epsfxsize=12cm   
\centerline{\epsfxsize=4.5in\epsfbox{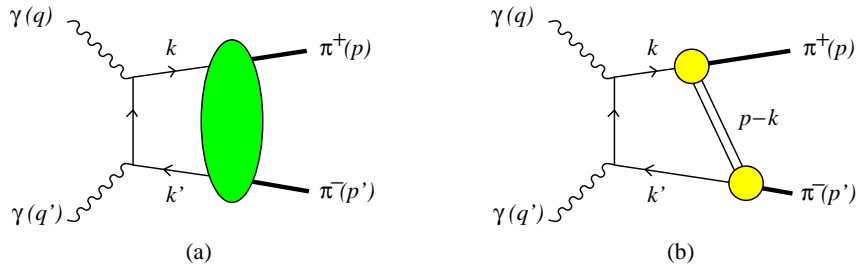}}   
\caption{(a) Handbag factorization of the process
$\gamma\gamma\to\pi\pi$ for large $s$ and $t$. The hard scattering
subprocess is shown at leading order in $\alpha_s$, and the blob
represents the two-pion distribution amplitude. The second
contributing graph is obtained by interchanging the photon vertices.
(b) The handbag resolved into two pion-parton vertices connected by
soft partons.  There is another diagram with the $\pi^+$ and $\pi^-$
interchanged.}
\label{fig:sketch}
\end{figure}

In the following, we discuss a complementary approach~\cite{DKV2} 
for large values of $s,\, t, \, u$, where the process 
amplitude factorizes into a hard subprocess for the production 
of a single $q\bar{q}$ pair, and a subsequent soft transition
$q\bar{q} \to \pi \pi$ (see Fig.\ref{fig:sketch}). 
The latter is described in terms of a new annihilation form 
factor which is given by a moment of the 
two-pion distribution amplitude~\cite{mue1994,die1998a}. 
This mechanism is analogous to the handbag contribution to 
wide-angle Compton scattering~\cite{rad1998,DFJK1}.

\section{The Physics of the Handbag Mechanism}

We consider the process $\gamma\gamma \to \pi^+\pi^-$ in the 
kinematical region where $s \sim -t \sim -u$. The condition 
for the transition $q\bar{q} \to \pi^+\pi^-$ to be soft is that
there be no large invariants at the parton-hadron vertices, cf. 
Fig.~\ref{fig:sketch}b. In particular, all virtualities occurring 
at these vertices are to be of $O(\Lambda^2)$, where 
$\Lambda$ is a typical hadronic scale. 
Moreover, the momenta of the additional $q\bar{q}$ pair 
and possibly other partons are required to be soft. This implies
that up to corrections of $O(\Lambda^2/s)$ the initial quark
and antiquark approximately carry the momenta of their respective
parent pions, and we have the condition $k \simeq p$ or $k' \simeq p$.

Note that configurations where the blob in Fig.~\ref{fig:sketch}a
contains hard gluon exchange are part of the leading-twist 
contribution and not included in the the soft handbag amplitude.
There are also diagrams of the cat's-ears topology, where the photons 
couple to different quark lines. At large $s$, $t$, $u$, these 
diagrams however require the presence of a large virtuality
in one of the quark lines or hard gluon exchange. 

In order to display the factorization it is advantageous to 
choose a symmetrical c.m. frame where the pions carry the same
light-cone plus momentum, i.e., the skewness, defined by
$\zeta=p^+/(p+p')^+$, has the value $1/2$. 
From the collinearity condition it follows that the initial 
quark and antiquark also have approximately equal light-cone
plus momenta. Thus, we have 
$z=k^+/(p+p')^+ = 1/2 + O(\Lambda^2/s)$.
Furthermore, corrections from partonic off-shell effects and
partonic transverse momenta are of $O(\Lambda^2/s)$.
Exploiting the on-shell and collinearity conditions one can 
show that the helicity amplitude for the process 
$\gamma \gamma \to \pi^+\pi^-$ can be written in the simple
form 
\begin{equation}
 A_{\mu\mu'} = - 4 \pi \ale \, \delta_{\mu,-\mu'} 
                      \, \frac{s^2}{t u} \, R_{2\pi}(s) \,,
\end{equation} 
where $\mu, \, \mu'$ are the photon helicities and the soft
part is encoded in the annihilation form factor defined by
\begin{equation}
   R_{2\pi}(s) = \sum_q e_q^2 \, R^q_{2\pi}(s) \,, \quad 
   R^q_{2\pi}(s) = \frac{1}{2} \int_0^1 d z \, (2\, z-1) \, 
        \Phi_{2\pi}^{q}(z,1/2,s) \,.
\end{equation} 
Here the summation is over $u, \, d, \, s$ quarks and  
$\Phi_{2\pi}^{q}(z,1/2,s)$ is the two-pion distribution 
amplitude in light-cone gauge~\cite{die1998a} at $\zeta = 1/2$.
We remark that the operator corresponding to this form factor 
is the quark part of the energy-momentum tensor, and that the
form factor is $C$ even due to the weight $(2 \, z -1)$, as one
could expect for a pion pair produced in two-photon annihilation.
The differential cross section of the process is given by
\begin{equation}
 \frac{d\sigma}{d t}(\gamma\gamma \to \pi^+\pi^-) = 
    \frac{8 \pi \ale^2}{s^2} \, \frac{1}{\sin^4\theta}\;
    \Big|R_{2\pi}(s)\Big|^{\,2} \,.
\label{dsdt-pipi}
\end{equation}

\section{Properties and Predictions}

Considerations of isospin of the intermediate $q\bar{q}$ pair
together with charge conjugation of the final state imply that 
charged and neutral
pion pairs are only produced in isospin zero states. Hence,
one can relate their corresponding form factors and finds the
key result  
\begin{equation}
 \frac{d\sigma}{dt}(\gamma\gamma \to \pi^0\pi^0) = 
       \, \frac{d\sigma}{dt}(\gamma\gamma \to \pi^+\pi^-)  \,,
\end{equation}
which is a parameter-free prediction of the handbag approach
and in striking contrast to the leading-twist approach, where 
the differential cross sections for $\pi^0 \pi^0$ production 
is found about an order of magnitude smaller than that for 
$\pi^+ \pi^-$ pairs.
By using $U$-spin symmetry, i.e., the symmetry under the exchange
$d \leftrightarrow s$, one can further relate the form factor
for $K^+ K^-$ to that of $\pi^+ \pi^-$ production, which leads
to the relation
\begin{equation}
  \frac{d\sigma}{dt}(\gamma\gamma \to K^+ K^-) \simeq 
        \, \frac{d\sigma}{dt}(\gamma\gamma \to \pi^+\pi^-) \,. 
\label{kaons}
\end{equation}
The approximate symbol indicates that, in general, flavour 
symmetry breaking effects have to be expected. Note 
that~(\ref{kaons}) holds in any dynamical approach respecting
SU(3) flavour symmetry. Isospin also provides a link between 
the form factors for charged and neutral kaon pairs, resulting in 
a further relation characteristic for the handbag mechanism:
\begin{equation}
\frac{d\sigma}{dt}(\gamma\gamma\to K^0\overline{K}{}^0) 
  \simeq \frac{4}{25} \,
                  \frac{d\sigma}{dt}(\gamma\gamma \to K^+K^-)\,,
\end{equation}
where we have neglected non-valence contributions and the numerical 
factor stems from the ratio of the corresponding charge factors.
 
\begin{figure}[t]
\centerline{\epsfxsize=2.2in \,\, \epsfbox{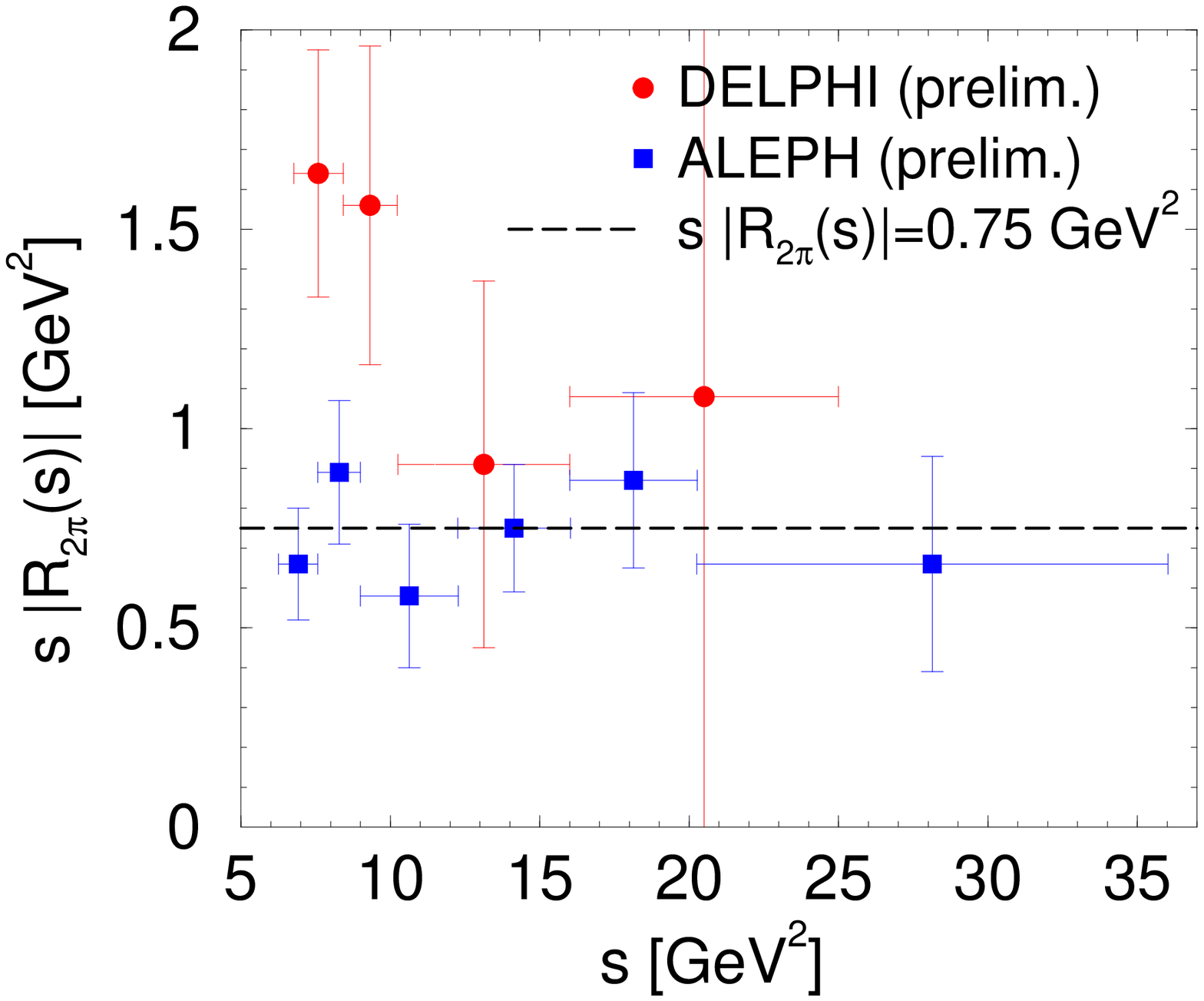}
            \epsfxsize=2.2in\epsfbox{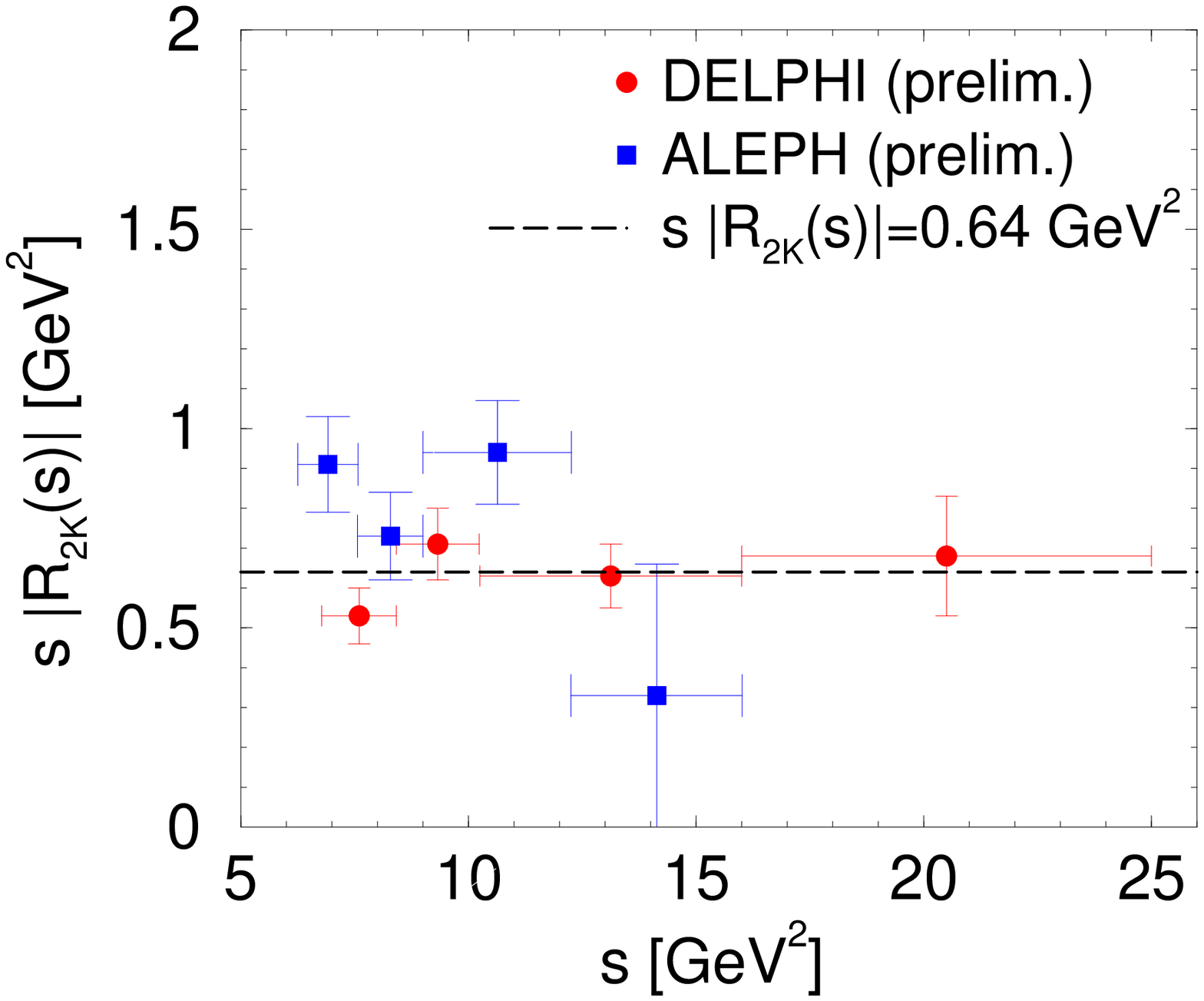}}  
\caption{The scaled annihilation form factors $s|R_{2\pi}|$ (left) and
$s|R_{2K}|$ (right) versus~$s$. The preliminary ALEPH and DELPHI data
is taken from \protect\cite{aleph,delphi}. Dashed lines represent our 
fitted values (\protect\ref{eq:num-ff}).}
\label{fig:formfactors}
\end{figure}

The annihilation form factors and the two-pion distribution 
amplitude can as yet not be calculated within QCD. They also 
do not allow for an overlap representation of light-cone wave
functions~\cite{DFJK3}. Therefore, we use the preliminary data
of ALEPH~\cite{aleph} and DELPHI~\cite{delphi} on the new
measurements of $\gamma \gamma \to \pi^+ \pi^-, \, K^+ K^-$ 
in order to extract the form factors from experiment. We find
that the scaling of the form factors is compatible with dimensional
counting rule behaviour and a fit for $\sqrt{s} \gsim 2.5 \gev$ 
provides the values (cf.~Fig.~\ref{fig:formfactors})
\begin{equation}
 s |R_{2\pi}(s)| = 0.75 \pm 0.07\gev^2 \quad \textrm{and} \quad
 s |R_{2K}(s)| = 0.64 \pm   0.04 \gev^2 \,.
\label{eq:num-ff}
\end{equation}
The pion annihilation form factor is comparable in magnitude 
with the timelike electromagnetic pion form factor.
Note that although the handbag contribution formally provides a
power correction to the leading-twist contribution, it appears
to dominate at experimentally accessible energies.

\begin{figure}[t]
\centerline{\epsfxsize=2.1in\epsfbox{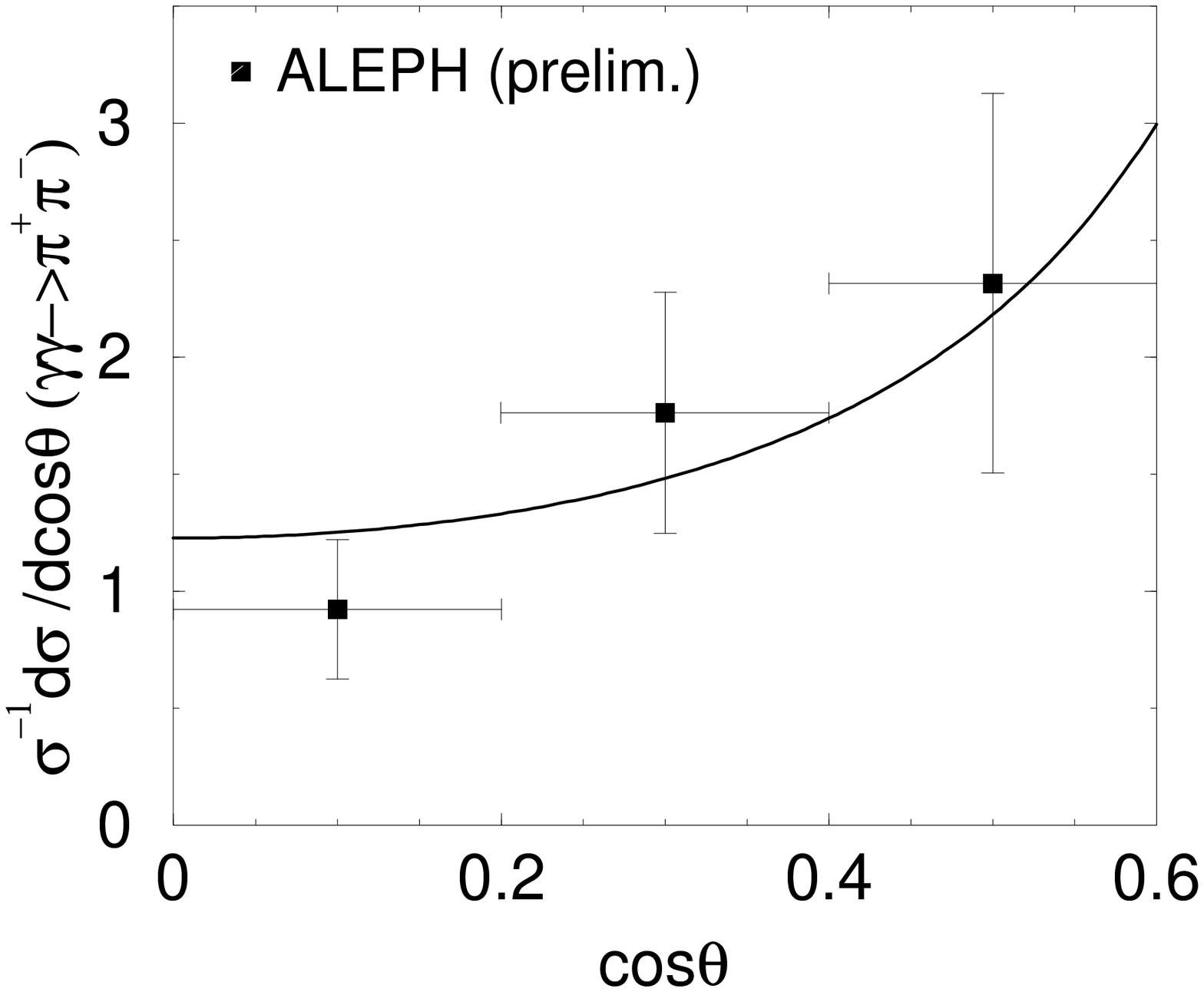}
            \epsfxsize=2.1in\epsfbox{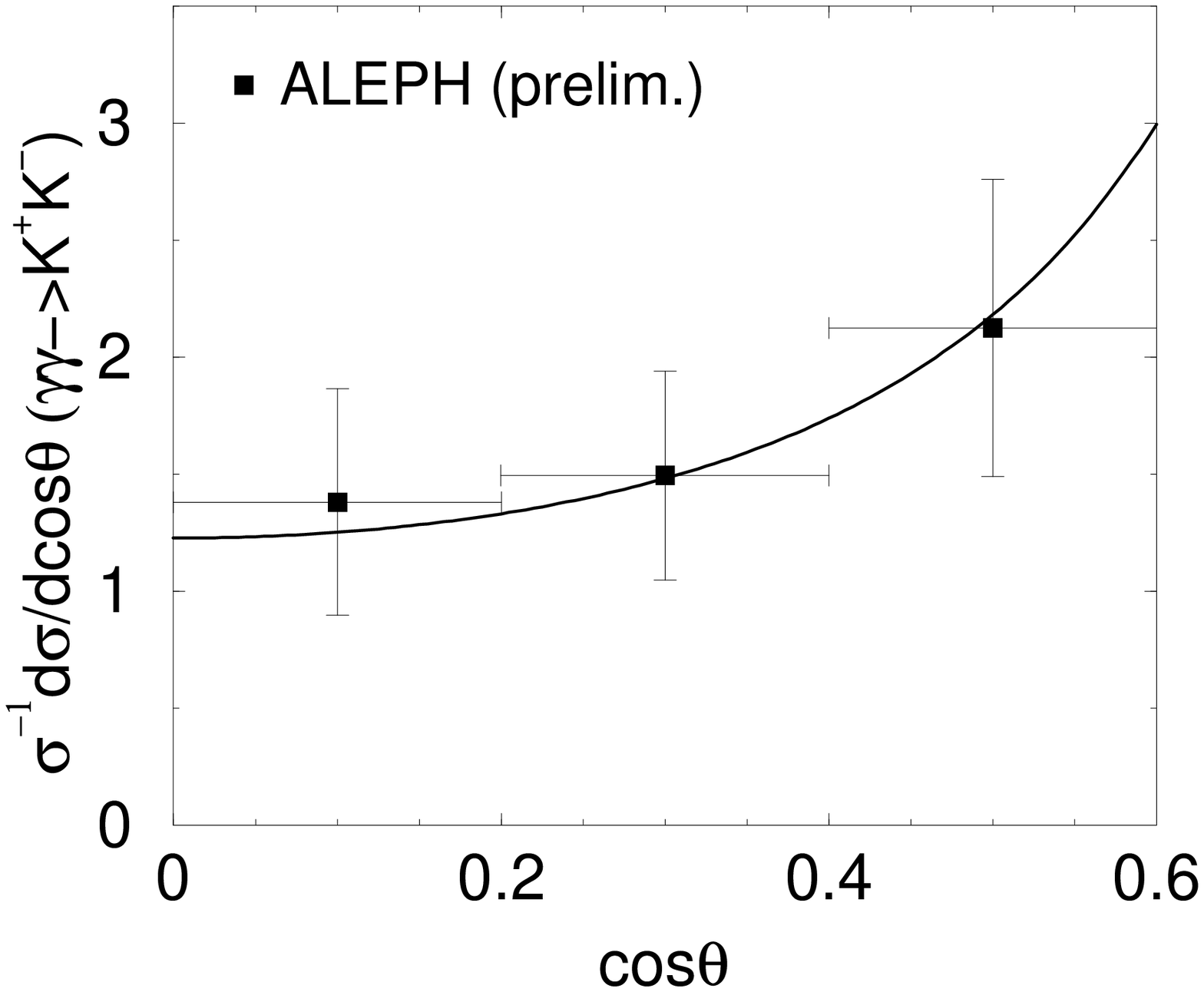}}  
\caption{The normalized angular distribution for 
$\gamma\gamma\to\pi^+\pi^-$ (left) and 
$\gamma\gamma\to K^+K^-$ (right), compared to the preliminary
ALEPH data~\protect\cite{aleph} for $4 \gev^2 <s <36 \gev^2$.}
\label{fig:angdis}
\end{figure}

Another characteristic result of our approach is the angular 
dependence (see Eq.~(\ref{dsdt-pipi})), which is in good 
agreement with the preliminary ALEPH data, as can be seen 
from Fig.~\ref{fig:angdis}. 

\begin{figure}[t]
\centerline{\epsfxsize=2.3in\epsfbox{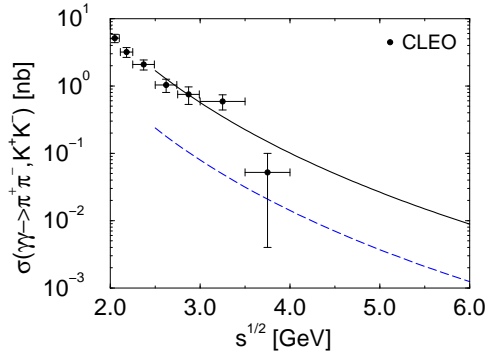}}
\caption{The CLEO data \protect\cite{cleo} for the cross section
$\sigma(\gamma\gamma\to \pi^+\pi^-) + \sigma(\gamma\gamma\to K^+K^-)$
integrated with $|\cos{\theta}| < 0.6$.  The solid line is the result
of the handbag approach with our fitted annihilation form factors
(\protect\ref{eq:num-ff}). The dashed line is the estimate of the 
leading-twist contribution described in the text.}
\label{fig:cleo}
\end{figure}

Having determined the normalization of the pion and kaon 
annihilation form factors from experiment, we can also compare
with the CLEO data for the integrated cross section~\cite{cleo}, 
where pions and kaons have not been separated. The result in 
displayed in Fig.~\ref{fig:cleo} and again we find good agreement.

As already mentioned in the introduction, the perturbative
contribution is way below the experimental data~\cite{vogt2000}. 
In order to facilitate comparison of the handbag and the 
leading-twist approach, we make a rather conservative estimate 
of the latter and employ a fixed coupling $\als=0.5$. We further 
use the asymptotic form for both the pion and kaon distribution 
amplitudes.
The leading-twist prediction thus obtained amounts to about 15\%
of our fitted handbag result as shown in Fig.~\ref{fig:cleo}.
%

\section{Conclusions}

We have presented a brief discussion of the handbag contribution 
to $\gamma \gamma \to \pi \pi, \, KK$. In this approach the 
process amplitude factorizes into a hard $\gamma \gamma \to q\bar{q}$
subamplitude and an annihilation form factor for the soft
transition to a meson pair. The form factor is a moment of the 
two-meson distribution amplitude at skewness $\zeta=1/2$. 
In lack of a model for the form factors and the two-meson
distribution amplitude in the kinematical range we are interested, 
we fit the form factors for $\pi^+ \pi^-$ and 
$K^+ K^-$ to the data and we find that their scaling is compatible 
with dimensional counting rule behaviour. Note that in the 
spacelike case of wide-angle Compton scattering off protons
the overlap representation in terms of light-cone wave functions,
together with a plausible model for the latter, explicitly shows 
how the soft part of the Compton form factors can mimic counting 
rule behaviour. 

Key results of the
approach are the prediction that the differential cross sections
for $\pi^+ \pi^-$ and $\pi^0 \pi^0$ production are the same, and 
the angular distribution $d\sigma/(d\cos\theta) \propto 1/ \sin^4\theta$, 
which agrees well with data. 

The handbag mechanism has recently 
also been applied the production of baryon-antibaryon 
pairs~\cite{DKV3}. In Ref.~\cite{weiss:2002} $p\bar{p}$ 
annihilation into photon pairs is investigated in an 
approach based on double distributions.

\section*{Acknowledgments}
I would like to thank M.~Diehl and J.~T.~Lenaghan for a 
careful reading of the manuscript.
This work is partially funded by the European
Commission IHP program under contract HPRN-CT-2000-00130.

\end{document}